\documentclass[11pt, onecolumn]{article} 
\usepackage{times,bm}
\usepackage[affil-it,]{authblk}
\usepackage[pagebackref=false,colorlinks=true,linkcolor=acsblue,citecolor=red,urlcolor=acsblue]{hyperref}

\usepackage{geometry}
\usepackage[switch*, displaymath,pagewise, mathlines]{lineno}
\usepackage{graphicx}
\usepackage{cite}
\usepackage{amsmath}
\usepackage{amsfonts}
\usepackage{amssymb}
\usepackage{slashed}
\usepackage{subfloat}
\usepackage{slashed}
\usepackage{color}
\usepackage{float}
\usepackage{multirow,multicol}
\usepackage{tabulary}
\usepackage[flushleft]{threeparttable}
\usepackage{pgfplots,subfigure}
\usepackage{xcolor}
\usepackage{caption} 
\numberwithin{equation}{section}
\usepackage{tikz}
\usepackage{ulem}
\usepackage{hyperref}
\usepackage{nameref}
\usepackage{comment}

\usepgflibrary{arrows}
\usetikzlibrary{shapes.callouts}
\tikzset{  
	level/.style   = { thick, },
	connect/.style = { dotted, red   },
	notice/.style  = { draw, rectangle callout, callout relative pointer={#1} },
	label/.style   = { text width=2cm }
}
\definecolor{acsblue}{RGB}{17,76,139}
\definecolor{shadecolor}{RGB}{255,241,204}
\usepackage{letltxmacro}
\makeatletter
\let\oldr@@t\r@@t
\def\r@@t#1#2{%
	\setbox0=\hbox{$\oldr@@t#1{#2\,}$}\dimen0=\ht0
	\advance\dimen0-0.2\ht0
	\setbox2=\hbox{\vrule height\ht0 depth -\dimen0}%
	{\box0\lower0.4pt\box2}}
\LetLtxMacro{\oldsqrt}{\sqrt}
\renewcommand*{\sqrt}[2][\ ]{\oldsqrt[#1]{#2}}
\makeatother
\usepackage{environ}
\begin{document}
\newcommand{{\ri}}{{\rm{i}}}
\newcommand{{\Psibar}}{{\bar{\Psi}}}

\title{\mdseries{Fermion-antifermion pairs in Bonnor-Melvin magnetic space-time with non-zero cosmological constant}}
\author{ \textit{Abdullah Guvendi}$^{\ 1}$\footnote{\textit{E-mail: abdullah.guvendi@erzurum.edu.tr (Corr. Auth.) } }~,~ \textit {Omar Mustafa}$^{\ 2}$\footnote{\textit{E-mail: omar.mustafa@emu.edu.tr} }  \\
	\small \textit {$^{\ 1}$ Department of Basic Sciences, Faculty of Science, Erzurum Technical University, 25050, Erzurum, Turkey}\\
	\small \textit {$^{\ 2}$ Department of Physics, Eastern Mediterranean University, G. Magusa, north Cyprus, Mersin 10 - Turkey}\\}

%\textbf{\Large ArXiv ePrint: \textcolor{acsblue}{1234.56789}}\\

\date{}
\maketitle
\begin{abstract}
We study the relativistic dynamics of fermion-antifermion pairs in the Bonnor-Melvin magnetic (BMM) spacetime in non-zero cosmology. We focus on the $(1+2)$-dimensional cylindrically symmetric BMM-spacetime background. Within the context of such a magnetized universe, we rigorously investigate the fully-covariant two-body Dirac equation. We derive the corresponding radial part of the equation and obtain an exact closed form analytical solution for the problem at hand. Moreover, we report some intriguing parallels with relativistic and non-relativistic quantum oscillators in flat spaces. Notably, our findings suggest a compelling correlation between the cosmological constant and the energy spectrum of fermion-antifermion systems, hinting at profound connections between quantum realms and cosmology.
 
\end{abstract}

\begin{small}
\begin{center}
\end{center}
\end{small}

%\begin{small}
%PACS numbers: 02.40.-k; 03.65.Ge; 03.65.-w; 04.20.Gz; 04.20.Jb; 04.62.+v;21.45.+v.
%\end{small}

\bigskip

\section{\mdseries{Introduction}}\label{sec1}

In the pursuit of understanding the dynamics of interacting fermions within the framework of relativistic quantum mechanics, the investigation involves the exploration of many-body equations constructed phenomenologically. These equations typically encompass the free Dirac Hamiltonian for each fermion along with interaction potentials. However, a notable challenge arises from relative time problems stemming from retardation effects, often disregarded in these phenomenological equations, despite their inherent involvement in a many-time problem. Complications also arise in selecting interaction potentials and determining the total angular momentum of fermion-formed systems within these equations. Conventionally, interaction potentials are simplified as either a single boson or photon exchange potential. Moreover, these phenomenologically established many-body equations fail to maintain full covariance in curved spaces. Contrasting this approach, non-relativistic quantum mechanics traditionally relies on one-time equations comprising individual particle free Hamiltonians and an interaction term, with wave functions dependent on spatial coordinates for each particle. The introduction of the Dirac equation prompted research into formulating a comprehensive two-body Dirac equation, notably initiated by Breit \cite{breit}, incorporating two free Dirac Hamiltonians and a modified interaction potential, akin to the Darwin potential in electrodynamics. However, limitations arise under conditions of long-range interactions or high particle velocities due to retardation effects. Subsequent attempts by Bethe and Salpeter, leveraging quantum field theory, encountered relative time problems, necessitating approximations like instantaneous interactions to solve for interacting fermions \cite{bs}. Decades later, Barut proposed a fully covariant many-body Dirac equation encompassing spin algebra and the most general electric and magnetic potentials \cite{barut}. Barut's equation, characterized by a spin algebra involving Kronecker products of Dirac matrices, results in a $16\times16$ dimensional matrix equation in $(3+1)$-dimensions. Achieving separation between angular and radial components involves employing group theoretical methods. Nevertheless, solving the subsequent set of $16$ radial equations remains an unsolved challenge even for familiar systems like one-electron atoms or unstable systems such as positronium. The challenge arises due to the interconnection of radial equations, resulting in pairs of coupled second-order wave equations -a hurdle that demands attention. However, in the past few years, there's been evidence indicating that the Barut equation is entirely solvable when dealing with low-dimensional systems or systems exhibiting particular dynamical symmetries, irrespective of whether they're in flat or curved spaces. This equation has seen remarkable progress in its application, marking significant breakthroughs \cite{a1,a2,a3,a4}. Hereby, in the current methodical proposal, our investigation seeks to elucidate the behavior of fermion-antifermion pairs in a magnetized universe (magnetized Bonnor-Melvin universe), specifically addressing the impact of non-zero cosmological constant.

A notable example of a fascinating curved space arises from the Bonnor-Melvin universe, an exact solution within the framework of the Einstein-Maxwell equations. Initially describing a static, cylindrically symmetric (electro)magnetic field embedded in its gravitational counterpart \cite{r7,r8,r9}, this solution was extended to encompass a non-zero cosmological constant ($\Lambda$), resulting in a magnetized, static, and cylindrically symmetric universe model \cite{r10}. The inclusion of $\Lambda$ significantly influences the overall geometry of spacetime, altering the behavior of the universe on a large scale. In particular, it impacts the expansion or contraction dynamics and gravitational behavior within the cylindrical structure described by the Bonnor-Melvin solution. The interaction between the cosmological constant and magnetic fields, as depicted in this unique solution of general relativity, sheds light on fundamental aspects of physics within Einstein's gravity theory. Moreover, understanding the evolution of vector fields in the presence of a magnetic field is crucial for investigating phenomena like magnetohydrodynamics, especially relevant in astrophysical contexts, such as plasma behavior in space. Exploring such magnetized curved spacetime can offer insights into early universe magnetic fields and their evolutionary paths, providing a foundation for theories about their origin and development on cosmological scales.

On the other hand, the presence of cylindrical symmetry implies that quantum particle dynamics remain invariant under Lorentz boosts (rotation-free Lorentz transformations) in the $\hat{z}$, significantly simplifying the problem at hand. This simplification allows for a clearer comprehension of dynamics and often streamlines calculations \cite{r11,r12,r13,r14,r15}. Relativistic quantum systems exhibiting dynamical symmetry refer to those governed by quantum mechanics, displaying specific symmetries within the framework of special relativity. These symmetrical behaviors in quantum systems play a pivotal role in understanding their properties, often leading to equations or Hamiltonians possessing invariant properties under particular transformations. The concept of dynamical symmetry facilitates the mathematical description of these systems, aiding in solving intricate quantum problems and predicting observable quantities. This understanding not only identifies patterns and predicts behavior but also underpins various applications, ranging from fundamental physics research to technological advancements in fields like materials science and beyond \cite{a1,a3,r12,r13}.

In this study, our focus revolves around investigating the intricate dynamics of fermion-antifermion pairs within a magnetized universe embedded in three dimensions and characterized by a non-zero cosmological constant $\Lambda$. Building upon the foundational work delineated in \cite{r10}, which established a four-dimensional magnetic universe governed by $\Lambda$, our research delves into a space-time configuration manifesting cylindrical symmetry. Notably, this configuration preserves the quantum field dynamics' invariance under Lorentz boosts along the $z$-direction, thereby facilitating a streamlined examination in $(1+2)$-dimensions. Our approach commences by deducing the radial wave equation governing bi-local fields, subsequently yielding its analytical solution through the utilization of special functions. Our investigation underscores a pivotal revelation: the cosmological constant $\Lambda$ exerts a profound influence on the eigenvalue solutions, underscoring their significant role in shaping the characteristics of these fermion-antifermion systems.

This manuscript is organized into several key sections addressing the dynamics of fermion-antifermion systems within a magnetized universe under non-zero cosmological conditions. Section \ref{sec2} lays the foundation by introducing the covariant two-body Dirac equation, initiating the derivation process for a wave equation applicable to such systems. Following this, in section \ref{sec3}, a non-perturbative wave equation is presented, accompanied by analytical results pertinent to this specialized scenario. Subsequently, a comprehensive overview of the outcomes will be presented, followed by an in-depth analysis and exploration of the findings. Throughout this manuscript, we will use the natural units.

\section{\mdseries{Two-body Dirac equation}} \label{sec2}

Within this section of the manuscript, we commence by presenting the fundamental structure of the covariant two-body Dirac equation in a $(2+1)$- dimensional curved space. Subsequently, we derive the specific form of this equation for a fermion pair in a magnetized $(2+1)$-dimensional magnetized universe under non-zero cosmology. The generalized form of the covariant two-body equation in $(2+1)$-dimensions manifests as follows \cite{a1,a2,a3,a4}
\begin{eqnarray}
&\left[\mathcal{H}^{f}\otimes \gamma^{t^{\overline{f}}}+\gamma^{t^{f}}\otimes \mathcal{H}^{\overline{f}} \right] \Psi(x^{\mu^{f}},x^{\mu^{\overline{f}}})=0,\nonumber\\
&\mathcal{H}^{f}=\left\lbrace \gamma^{\mu^{f}}\slashed{\mathcal{D}}_{\mu}^{f}+im\textbf{I}_2 \right\rbrace, \mathcal{H}^{\overline{f}}=\left\lbrace \gamma^{\mu^{\overline{f}}}\slashed{\mathcal{D}}_{\mu}^{\overline{f}}+im\textbf{I}_2 \right\rbrace,\nonumber\\
&\slashed{\mathcal{D}}_{\mu}^{f}=\partial_{\mu}^{f}-\Gamma^{f}_{\mu},\quad \slashed{\mathcal{D}}_{\mu}^{\overline{f}}=\partial_{\mu}^{\overline{f}}-\Gamma^{\overline{f}}_{\mu},\label{eq1}
\end{eqnarray}
in which $x^{\mu}$ represents the coordinates within this curved space, $f\, (\overline{f})$ denotes fermions (antifermions), $\gamma^{\mu}$ stand as space-dependent Dirac matrices, $m$ signifies the rest mass of individual particles, $\Psi(x^{\mu^{f}},x^{\mu^{\overline{f}}})$ signifies the bi-local spinor field that depends on the spacetime position vectors ($x^{\mu^{f}},x^{\mu^{\overline{f}}}$) of the fermions, $\textbf{I}_2$ denotes the 2-dimensional identity matrix, and $\Gamma{\mu}$ represents the spinorial affine connections for each Dirac field. The magnetic universe (with a non-zero cosmological constant) can be described by the following line element 
\begin{eqnarray}
&ds^{2}=g_{\mu\nu}dx^{\mu}dx^{\nu}=dt^2-dr^2-\diamondsuit(r)^2 d\phi^2,\nonumber\\
&\diamondsuit(r)=\sigma\, \sin\left(\sqrt{2\Lambda}\,r\right).\label{eq2}
\end{eqnarray}
The relationship between $\sigma$ (an integration constant) and $\Lambda$ is connected to the magnetic field. This magnetic field is defined by $H=\sqrt{\Lambda}\sigma \sin(\sqrt{2\Lambda}r)$, as found in \cite{r10}. In this context, Greek indices are used to denote coordinates within the curved space-time, represented as $x^{\mu}=t, r, \phi$. The Gaussian curvature ($\mathcal{K}$) of this particular space-time background can be calculated as $\mathcal{K}=-\frac{\diamondsuit_{,rr}}{\diamondsuit}=2\Lambda$, where the subscript $_{,r}$ signifies derivative with respect to $r$. Next, the derivation of the space-dependent Dirac matrices involves employing the expression: $\gamma^{\mu}=e^{\mu}_{k}\gamma^{k}, (k=0,1,2.)$. The quantities $e^{\mu}_{k}$ correspond to the inverse tetrad fields, while $\gamma^{k}$ symbolize the space-independent Dirac matrices. The Latin indices within this context denote the coordinates within a flat spacetime. The selection of $\gamma^{k}$ relies on the familiar Pauli matrices ($\sigma^{x}, \sigma^{y}, \sigma^{z}$), determined by the signature ($+,-,-$) of the specific line element. This ensures $\gamma^{0}=\sigma^{z}$, $\gamma^{1}=i\sigma^{x}$, and $\gamma^{2}=i\sigma^{y}$, where $i=\sqrt{-1}$ \cite{a2,a3}. The determination of tetrad fields involves utilizing the expression $g_{\mu\nu}=e^{k}_{\mu}e^{l}_{ \nu}\tilde{\eta}_{kl}$ \cite{a2,a3}. Here, $\tilde{\eta}_{kl}$ signifies the flat Minkowski tensor characterized by a signature of $(+,-,-)$. The covariant metric tensor $g_{\mu\nu}$ is $g_{\mu\nu}=\textrm{diag}(1, -1, -\diamondsuit^{2})$ and its inverse becomes $g^{\mu\nu}=\textrm{diag}(1, -1, -\diamondsuit^{-2})$. The determination of the inverse tetrad fields ($e^{\mu}_{k}$) is feasible through the relationship $e^{\mu}_{k}=g^{\mu\nu}e^{l}_{\nu}\tilde{\eta}_{kl}$ in which $e^{l}_{\nu}$ are tetrad fields that can be determined through $g_{\mu\nu}=e^{k}_{\mu}e^{l}_{\nu}\tilde{\eta}_{kl}$. Accordingly, we have
\begin{eqnarray}
&e^{0}_{t}=1,\quad e^{1}_{r}=1,\quad e^{2}_{ \phi}=\diamondsuit,\nonumber\\
&e^{t}_{ 0}=1,\quad e^{r}_{ 1}=1, \quad e^{\phi}_{ 2}=\diamondsuit^{-1}. \label{eq3}
\end{eqnarray}
Here, it can be seen that the tetrad fields satisfy the orthogonality and orthonormality conditions. Also, now, we need to obtain the affine spin connections by using the relation: $\Gamma_{\eta}=\frac{1}{8}g_{\mu\lambda}\left(e^{q}_{\nu,\eta}e^{\lambda}_{q}-\Gamma_{\nu\eta}^{\lambda} \right)\left[\gamma^{\mu}\gamma^{\nu}-\gamma^{\nu}\gamma^{\mu}\right]$ where $\Gamma_{\nu\eta}^{\lambda}$ symbols can be calculated by using the expression given by $\Gamma_{\nu\eta}^{\lambda}=\frac{1}{2}g^{\lambda\epsilon}\left\lbrace\partial_{\nu}g_{\eta\lambda}+\partial_{\eta}g_{\epsilon\nu}-\partial_{\epsilon}g_{\nu\eta} \right\rbrace$ \cite{a2,a3}. Thereby, one can find that 
\begin{eqnarray}
\Gamma^{r}_{\phi\phi}=-\diamondsuit\diamondsuit_{,r}, \quad \Gamma^{\phi}_{r \phi}=\frac{\diamondsuit_{,r}}{\diamondsuit}.\nonumber
\end{eqnarray}
Finally, the outcomes regarding the generalized Dirac matrices and the non-vanishing component of the affine spin connections can be summarized as follows\footnote{Here, one should note that $\gamma^{t}$ and $ \gamma^{r}$ are independent from the $r$, but the others must be considered according to radial coordinates of $f$ and $\overline{f}$.}
\begin{eqnarray}
&\gamma^{t}=\sigma^{z},\quad \gamma^{r}=i\sigma^{x},\quad \gamma^{\phi}=\frac{i}{\diamondsuit}\sigma^{y},\nonumber\\
&\Gamma_{\phi}=\frac{i}{2}\diamondsuit_{,r}\sigma^{z}.\label{eq4}
\end{eqnarray}
It might be advantageous to express the covariant two-body equation explicitly, taking the form of $\hat{\Upsilon}\Psi=0$ in which $\hat{\Upsilon}$ is
\begin{eqnarray}
&\gamma^{t^{f}}\otimes\gamma^{t^{\overline{f}}}\left\lbrace \partial_{t}^{f}+\partial_{t}^{\overline{f}} \right\rbrace +\gamma^{r^{f}}\otimes \gamma^{t^{\overline{f}}}\partial_{r}^{f}+ \gamma^{t^{f}}\otimes \gamma^{r^{\overline{f}}}\partial_{r}^{\overline{f}}\nonumber\\
&+\gamma^{\phi^{f}} \otimes\gamma^{t^{\overline{f}}}\partial_{\phi}^{f} +\gamma^{t^{f}}\otimes \gamma^{\phi^{\overline{f}}}\partial_{\phi}^{\overline{f}}\nonumber\\
&+im\left\lbrace\textbf{I}_{2}\otimes \gamma^{t^{\overline{f}}}+\gamma^{t^{f}}\otimes \textbf{I}_{2}\right\rbrace\nonumber\\
&-\left\lbrace \gamma^{\phi^{f}}\Gamma_{\phi}^{f}\otimes \gamma^{t^{\overline{f}}}+\gamma^{t^{f}}\otimes \gamma^{\phi^{\overline{f}}}\Gamma_{\phi}^{\overline{f}} \right\rbrace ,   \label{eq5}
\end{eqnarray}
and it is obvious in the Eq. (\ref{eq4}) that $\gamma^{\phi}\Gamma_{\phi}=-\frac{i}{2}\frac{\diamondsuit_{,r}}{\diamondsuit} \sigma^{x}$. It's pertinent to mention that our aim is to analyze the relative motion of the considered pair in the magnetized universe affected by non-zero cosmological constant $\Lambda$. To accomplish this, separating the center of mass motion coordinates ($R$) and relative motion coordinates ($r$) becomes imperative, the process facilitated by the expressions elucidated in the following \cite{a1}:
\begin{eqnarray}
&r_{\mu}=x_{\mu}^{f}-x_{\mu}^{\overline{f}},\quad R_{\mu}=\frac{x_{\mu}^{f}}{2}+\frac{x_{\mu}^{\overline{f}}}{2}, \quad x_{\mu}^{f}=R_{\mu}+\frac{r_{\mu}}{2},\nonumber\\
& x_{\mu}^{\overline{f}}=R_{\mu}-\frac{r_{\mu}}{2},\quad \partial_{x_{\mu}}^{f}=\frac{\partial_{R_{\mu}}}{2}+\partial_{r_{\mu}},\nonumber\\
&\partial_{x_{\mu}}^{\overline{f}}=\frac{\partial_{R_{\mu}}}{2}-\partial_{r_{\mu}},\label{eq6}
\end{eqnarray}
and thus $\partial_{x_{\mu}}^{f}+\partial_{x_{\mu}}^{\overline{f}}=\partial_{R_{\mu}}$. Now, it is crucial to emphasize that the system's energy relates to the time coordinate of the center of mass motion (see Eq. (\ref{eq5})). This association arises due to the presence of the $\partial_{t}^{f}+\partial_{t}^{\overline{f}}$ term, resulting in $\partial_{t}^{f}+\partial_{t}^{\overline{f}}=\partial_{R_{t}}$. Assuming the center of mass for the pair remains stationary at the spatial origin, we can effectively decompose the spinor $\Psi$, $\Psi\left(t,r, \phi\right)=\textrm{e}^{-i\omega t}\textrm{e}^{is\phi}\left(\psi_{1}\, \psi_{2}\, \psi_{3}\, \psi_{4} \right)^{T}$, with $T$ representing the spinor's transpose depending on the relative radial coordinate $r$, and where $s$ signifies the combined spin of two fermions. Consequently, following some algebraic arrangements, one can deduce the subsequent equation(s)
\begin{flalign}
\omega \chi_{1}-\tilde{m}\chi_{2}-\frac{2s}{\sigma \sin\left(a r\right)}\chi_{3}+2\left[\partial_{r}+a\cot\left(ar\right)\right]\chi_{4}=0,\label{WE}
\end{flalign}
where
\begin{flalign*}
\chi_{2}=\frac{\tilde{m}}{\omega}\chi_{1},\quad \chi_{3}=\frac{2\,s}{\omega\, \sigma\, \sin\left(ar\right)}\chi_{1},\quad \chi_{4}=\frac{2}{\omega}\left[\partial_{r}+a\,\cot\left(ar\right)\right]\chi_{1}
\end{flalign*}
provided
\begin{flalign*}
&\chi_{1}(r)=\psi_{1}(r)+\psi_{4}(r),\quad \chi_{2}(r)=\psi_{1}(r)-\psi_{4}(r),\nonumber\\
&\chi_{3}(r)=\psi_{2}(r)+\psi_{3}(r),\quad \chi_{4}(r)=\psi_{2}(r)-\psi_{3}(r),
\end{flalign*} 
and $\tilde{m}=2m$, $a=\sqrt{\Lambda/2}$.
 
\section{\mdseries{Analytical solutions for the wave equation}} \label{sec3}

Here, we present a non-perturbative wave equation governing a fermion-antifermion pair within a magnetized universe affected by non-zero cosmological constant. Our objective is to seek a closed form analytical solution for this specific wave equation. The Eq. (\ref{WE}) leads to the subsequent wave equation manifestation
\begin{eqnarray*}
\chi_{1_{,rr}}(r)+2a \cot\left(ar\right)\chi_{1_{,r}}(r)+\left[\frac{\epsilon}{4}-a^2-\frac{s^2}{\vartheta(r)^2}\right]\chi_{1}(r)=0,
\end{eqnarray*}
where $\epsilon=\omega^2-\tilde{m}^2$ and $\vartheta(r)=\sigma \sin\left(a r\right)$. This equation, with the change of variable $y=\left[ 1+\cos \left( ar\right) %
\right] /2$, would yield%
\begin{equation}
y\left( y-1\right) \chi _{1,yy}\left( y\right) +\left( 3y-\frac{3}{2}\right)
\chi _{1,y}\left( y\right) -\left( \frac{\epsilon }{4a^{2}}-1+\frac{s^{2}}{%
4a^{2}\sigma ^{2}y\left( y-1\right) }\right) \chi _{1}\left( y\right) =0.
\label{e1}
\end{equation}%
We may now use the substitution%
\begin{equation}
\chi _{1}\left( y\right) =y^{\beta /2}\left( y-1\right) ^{\gamma /2}\chi
\left( y\right) ,  \label{e2}
\end{equation}%
where,%
\begin{equation}
\beta =-\frac{1}{2}+\frac{1}{2}\sqrt{1+\frac{4s^{2}}{a^{2}\sigma ^{2}}}%
,\;\gamma =-\frac{1}{2}-\frac{1}{2}\sqrt{1+\frac{4s^{2}}{a^{2}\sigma ^{2}}}
\label{e3}
\end{equation}%
to obtain%
\begin{equation}
y\left( y-1\right) \chi _{,yy}\left( y\right) +\left( 2y+\gamma -\frac{1}{2}%
\right) \chi _{,y}\left( y\right) +\frac{1}{4}\left( 1-\frac{\epsilon }{a^{2}}%
\right) \chi \left( y\right) =0.  \label{e4}
\end{equation}%
Obviously, this equation is in the form of  the Gauss's hypergeometric differential equation \cite{2,Abramowitz}. 
that admits a solution in the form of%
\begin{eqnarray}
\chi \left( y\right)  &=&\mathcal{N}\,M\left( \left[ \frac{1}{2}+\frac{1}{2}%
\sqrt{\frac{\epsilon }{a^{2}}},\frac{1}{2}-\frac{1}{2}\sqrt{\frac{\epsilon }{%
a^{2}}}\right] ,\frac{1}{2}\sqrt{1+\frac{4s^{2}}{a^{2}\sigma ^{2}}}%
+1,y\right)   \notag \\
&=&\mathcal{N}\,M\left( \frac{1}{2}\pm \frac{1}{2}\sqrt{\frac{\epsilon }{%
a^{2}}},\frac{1}{2}\sqrt{1+\frac{4s^{2}}{a^{2}\sigma ^{2}}}+1,y\right) .
\label{e005}
\end{eqnarray}%
where $\mathcal{N}$ is the normalization constant (the determination of
which is far beyond the scope of the current study). Next, we have to
truncate the hypergeometric power series into a polynomial of order $n\geq 0$
(to secure finiteness and square integrability of the radial wave function)
by requiring that%
\begin{equation}
\frac{1}{2}\pm \frac{1}{2}\sqrt{\frac{\epsilon }{a^{2}}}=-n%
\Longleftrightarrow \epsilon =\omega ^{2}-\tilde{m}^{2}=a^{2}\left(
2n+1\right) ^{2}.  \label{e6}
\end{equation}%
Consequently, 
\begin{equation}
\omega _{n}=\pm \sqrt{\tilde{m}^{2}+a^{2}\left( 2n+1\right) ^{2}}%
;\,n=0,1,2,\cdots .  \label{e7}
\end{equation}

However, it could be interesting to know that a power series expansion in the
form of 
\begin{equation}
\chi \left( y\right) =\sum\limits_{j=0}^{\infty }C_{j}\,y^{j+\upsilon },
\label{e8}
\end{equation}%
in Eq. (\ref{e4}), would result%
\begin{equation}
\sum\limits_{j=0}^{\infty }C_{j}\left\{ \left[ \left( j+\upsilon \right)
\left( j+\upsilon +1\right) +\mathcal{E}\right] \,y^{j+\upsilon }+\left(
j+\upsilon \right) \left( \gamma +\frac{1}{2}-j-\upsilon \right)
y^{j+\upsilon -1}\right\} =0,  \label{e9}
\end{equation}%
where $\mathcal{E}=1/4-\epsilon /4a^{2}$. In a straightforward manner, one
obtains%
\begin{gather}
\sum\limits_{j=0}^{\infty }\left\{ C_{j}\left[ \left( j+\upsilon \right)
\left( j+\upsilon +1\right) +\mathcal{E}\right] \,+C_{j+1}\left( j+\upsilon
\right) \left( \gamma -\frac{1}{2}-j-\upsilon \right) \right\} y^{j+\upsilon
}  \notag \\
+C_{0}\left[ \upsilon \left( \gamma -\upsilon +\frac{1}{2}\right) \right]
\,y^{\upsilon -1}=0.  \label{e10}
\end{gather}%
Which, in turn, suggests a two terms recursion relation%
\begin{equation}
C_{j}\left[ \left( j+\upsilon \right) \left( j+\upsilon +1\right) +\mathcal{E%
}\right] \,+C_{j+1}\left( j+\upsilon \right) \left( \gamma -\frac{1}{2}%
-j-\upsilon \right) =0,  \label{e11}
\end{equation}%
and since $C_{0}\neq 0$  
\begin{equation}
C_{0}\neq 0\Longrightarrow \upsilon \left( \gamma -\upsilon +\frac{1}{2}%
\right) \Longrightarrow \upsilon =0,\;\upsilon =\gamma +\frac{1}{2}\text{.}
\label{e12}
\end{equation}%
It is clear that  $\upsilon =\gamma +\frac{1}{2}$ renders the radial wave
function infinite and unbounded at $y=0$ (i.e., at $\cos \left( ar\right) =-1
$). We adopt $\upsilon =0$, therefore. Under such settings, our two terms
recursion relation now reads%
\begin{equation}
C_{j}\left[ j\left( j+1\right) +\mathcal{E}\right] \,+C_{j+1}\left[ j\left(
\gamma -\frac{1}{2}-j\right) \right] =0.  \label{e13}
\end{equation}%
Our power series in Eq. (\ref{e8}) needs to be truncated into a polynomial of
order $n\geq 0$ by the requirement that $\forall j=n$ we set $C_{n+1}=0$.
Consequently, 
\begin{equation}
C_{n}\left[ n\left( n+1\right) +\mathcal{E}\right] \,=0\Longrightarrow
C_{n}\neq 0\Longrightarrow n\left( n+1\right) +\mathcal{E}=0.  \label{e14}
\end{equation}%
This result, along with  $\mathcal{E}=1/4-\epsilon /4a^{2}$ and $\epsilon
=\omega ^{2}-\tilde{m}^{2}$,  would imply 
\begin{equation}
\frac{1}{4}-\frac{\epsilon }{4a^{2}}=n\left( n+1\right) \Longleftrightarrow
\omega ^{2}-\tilde{m}^{2}=a^{2}\left( 2n+1\right) ^{2}.  \label{e15}
\end{equation}%
This result is in exact accord with that in (\ref{e6}) and implies that in (\ref{e7}).
%Let us attempt to rewrite this equation using a new variable change, $2y=1+\cos(ar)$, aiming to eliminate the trigonometric functions. By expressing this equation in terms of the variable $y$, we can effectively remove the trigonometric functions. Additionally, this enables us to observe the asymptotic tendencies of the resulting equation. Accordingly, by introducing an assumed function,
%\begin{eqnarray*}
%\chi_{1}(y)=y^{\xi/2}(y-1)^{\xi/2}\chi(y),
%\end{eqnarray*}
%in which $\xi={\frac{\sqrt{4s^2+a^2\sigma^2}}{2\sigma a}}-1/2$, the resulting equation becomes:
%\begin{flalign}
%&y\left(y-1\right)\chi_{,yy}-[\tilde{\gamma}-(\tilde{\alpha}+\tilde{\beta}+1)y]\chi_{,y}+\tilde{\alpha}\tilde{\beta}\chi=0,\nonumber\\ \label{eq9}, 
%\end{flalign}
%where $\tilde{\alpha}=-\frac{\sqrt{\epsilon}}{2a}+\frac{1}{2}$, $\tilde{\beta}=\frac{\sqrt{\epsilon}}{2a}+\frac{1}{2}$, $\tilde{\gamma}=1+\frac{\sqrt{4s^2+a^2\sigma^2}}{2a\sigma}$,
%$\tilde{\alpha}$, $\tilde{\beta}$, and $\tilde{\gamma}$ are constants. The Eq. (\ref{eq9}) is the Gauss's hypergeometric wave equation \cite{2,Abramowitz}. Its regular solution is denoted by $\chi\left(y \right)=\mathcal{Q}\, _{2}F_{1}\left(\tilde{\alpha}, \tilde{\beta} ; \tilde{\gamma}; y \right)$, where $\mathcal{Q}$ represents an arbitrary constant. This solution function takes the form of a polynomial of degree $n$ concerning the variable $y$ if $\tilde{\alpha}=-n$, with $n$ being the radial quantum number ($n=0,1,2..$). 
This condition leads to a specific frequency spectrum, which can be expressed as follows:
\begin{eqnarray}
\omega_{n}= \pm \left\lbrace \tilde{m}^2+4a^2\tilde{n}^{2}\right\rbrace^{1/2},\quad \tilde{n}=n+1/2.\label{FS}
\end{eqnarray}

Hereby, it is evident that the system's energy remains unaffected by the total spin of the fermion-antifermion pair, despite the explicit dependence of the wave function on the total spin. The form of the wave equation represented by Eq. (\ref{e4}) remains consistent even when $m=0$, enabling us to explore scenarios involving massless pairs. In situations where $m$ is set to zero, the frequency spectrum exhibits the pattern 
\begin{eqnarray}
\omega_{n}=\sqrt{2\Lambda}\tilde{n},\label{SO}
\end{eqnarray}
mirroring the characteristic findings of a one-dimensional non-relativistic quantum oscillator. It is worth noting that our findings bear a striking resemblance to the characteristics exhibited by a spinless relativistic quantum oscillator (specifically when $m\neq 0$) within flat spaces, while also mirroring the well-documented outcomes of a non-relativistic quantum oscillator (when $m=0$). The frequency spectrum described by Eq. (\ref{FS}) incorporates the spacetime parameter $\Lambda$, which is associated with the magnetic field, along with the quantum number $n$. Here, it is important to highlight the significant role of the cosmological constant in cosmology, especially concerning our comprehension of dark energy and its influence on the universe's evolution. The correlation between this constant and the energy spectrum of fermion-antifermion systems suggests a potential profound connection between the quantum domain and the wider scope of cosmology. Additionally, Eq. (\ref{FS}) resembles the precise outcomes obtained for relativistic oscillators (see also \cite{a1}). Within the realm of quantum field theory, vacuum fluctuations resemble an expansive array of harmonic oscillators spanning various frequencies. The energy stemming from these fluctuations adds to the vacuum energy, potentially associated (at a conceptual level) with the cosmological constant. Nonetheless, this correlation is intricate, forming a significant facet of continual theoretical exploration within quantum field theory, specifically in its implications for cosmology.

\begin{figure}[ht!]
   \centering
   \includegraphics[scale=0.60]{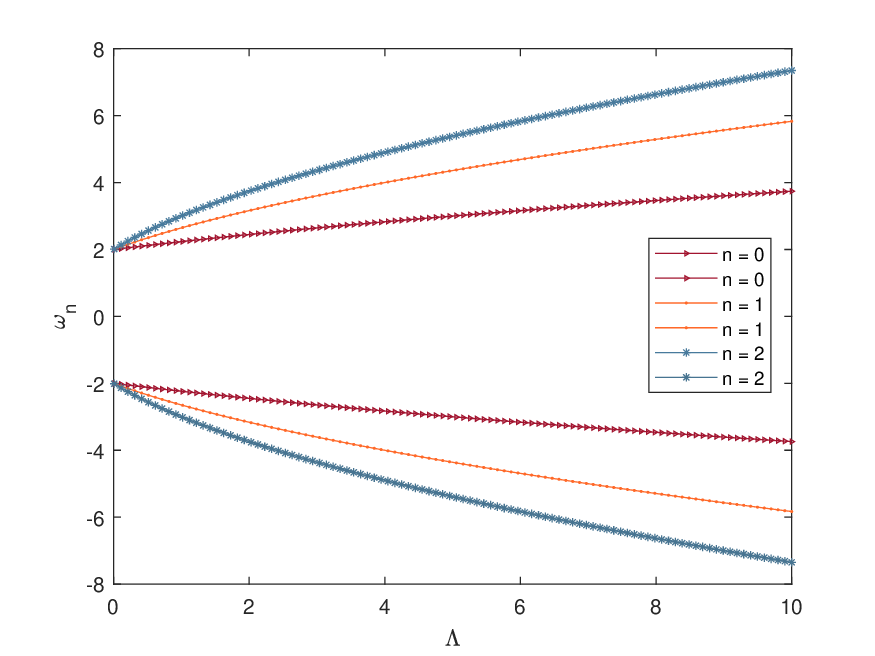}
    \caption{Relativistic frequencies for varying values of the $\Lambda$. Here, we take $m=1$.}
    \label{fig1}
\end{figure}

\section{\mdseries{Discussions}}\label{sec5}

In this study, our exploration of a specific curved geometry within three-dimensional space featuring a cosmological constant can be significantly contributed to the broader scope of research in relativistic particle dynamics. By focusing on the eigenvalue solutions for fermion-antifermion pairs in a magnetized Bonnor-Melvin universe, characterized by a non-zero cosmological constant, we have unveiled intriguing insights into the behavior of fermion-antifermion pairs within this unique space-time configuration. The investigation of (2+1)-dimensional metric has yielded surprising parallels between relativistic fermionic behavior in this specialized setting and solutions akin to the relativistic spinless oscillator in flat space (see also \cite{a1} and Figure \ref{fig1}). Notably, our precise computations have revealed compelling similarities between the dynamics of fermion-antifermion pairs in this background and those encountered in relativistic and non-relativistic oscillator frameworks (see Eq. (\ref{FS})). This unexpected correspondence not only sheds light on the intricate interplay between relativistic fermions and the distinct geometry of the Bonnor-Melvin universe but also suggests an intriguing connection between seemingly disparate physical phenomena. Moreover, our examination of massless fermion-antifermion pairs within this universe model has provided interesting results showing similarities to the behavior of a non-relativistic one-dimensional quantum oscillator (see Eq. (\ref{SO})). This significant finding holds implications for comprehending massless particles in various cosmological settings and urges further investigation into the fundamental principles governing their behavior in such intriguing space-time configurations. Our comprehensive analysis not only highlights the nuances of fermionic dynamics in magnetized Bonnor-Melvin universes with non-zero cosmological constants but also underscores unexpected connections between relativistic fermionic behavior and well-established phenomena in flat space. The existence of the cosmological constant $\Lambda$ in the obtained frequency spectra emphasizes a profound link between the system's energy and the fundamental nature of the universe's evolution. The presence of the cosmological constant within the frequency spectrum implies that space-time dynamics, influenced by the $\Lambda$, intricately impact the energy levels of fermion-antifermion pairs. This connection presents an opportunity to glean insights into the interplay between quantum mechanics and cosmology. The value of the cosmological constant significantly determines the universe's evolution rate and plays a pivotal role in cosmological models. Within the realm of quantum field theory, vacuum fluctuations, akin to an infinite collection of harmonic oscillators with different frequencies, contribute to the vacuum energy. Conceptually linking this energy to the cosmological constant remains a complex endeavor and is part of ongoing theoretical research in quantum field theory, offering tantalizing implications for our understanding of cosmological constants and their relation to fundamental physical phenomena.

%\section*{\small{Acknowledgements}}
%The author thanks the anonymous referee for careful reading, valuable comments and kind suggestions.\\

\section*{\small{Data availability}}
This manuscript has no associated data or the data will not be deposited.

\section*{\small{Conflicts of interest statement}}
No conflict of interest declared by the authors.

\section*{\small{Funding}}
No funding regarding this research.

\end{document}